\documentclass{aipproc}
\layoutstyle{6x9}
\usepackage{amssymb}

\newcommand{\sauron}{\texttt{SAURON}}
\newcommand{\oasis}{\texttt{OASIS}}
\newcommand{\mnras}{MNRAS}
\newcommand{\apj}{ApJ}
\newcommand{\apjl}{ApJ}
\newcommand{\aj}{AJ}
\newcommand{\aap}{A\&A}
\newcommand{\nat}{Nature}

\newcommand{\araa}{ARA\&A}

\begin{document}

\title{Testing Mass Determinations of Supermassive Black Holes via Stellar Kinematics}


\author{
	Michele Cappellari$^1$,
	Richard M.\ McDermid$^2$,
	R.\ Bacon$^3$,
	Roger L.\ Davies$^1$,
	P.\ T.\ de Zeeuw$^{4,5}$,
	Eric Emsellem$^{3,4}$,
	Jes\'us Falc\'on-Barroso$^6$,
	Davor Krajnovi{\'c}$^4$,
	Harald Kuntschner$^7$,
	Reynier F.\ Peletier$^8$,
	Marc Sarzi$^9$,
	Remco C. E. van den Bosch$^{10}$,
	Glenn van de Ven$^{11}$}
{address={
    $^1$University of Oxford, UK;
    $^2$Gemini Observatory, Hilo, USA;
    $^3$University of Lyon, France;
    $^4$ESO, Garching, Germany;
    $^5$Leiden University, The Netherlands;
    $^6$IAC, La Laguna, Spain;
    $^7$ST-ECF, ESO, Garching, Germany;
    $^8$University of Groningen, The Netherlands;
    $^9$University of Hertfordshire, UK;
    $^{10}$University of Texas, Austin, USA;
    $^{11}$MPIA, Heidelberg, Germany
}}

\begin{abstract}
We investigate the accuracy of mass determinations $M_{\rm BH}$ of supermassive black holes in galaxies using dynamical models of the stellar kinematics. We compare 10 of our $M_{\rm BH}$ measurements, using integral-field \oasis\ kinematics, to published values. For a sample of 25 galaxies we confront our new $M_{\rm BH}$ derived using two modeling methods on the same \oasis\ data.
\end{abstract}

\maketitle

\section{Introduction}

Scaling relations between global parameters of galaxies and their central supermassive black holes (BHs) \citep{kormendy95,magorrian98,gebhardt00bhs,ferrarese00,marconi03,haring04} are central to our current understanding of how galaxies assemble in the hierarchical merging paradigm \citep{silk98,dimatteo05}. Thousands of publications made use of a small collection of $\sim50$ BH mass determinations $M_{\rm BH}$ accumulated from a large combined effort over the past 15 years (see compilations in \cite{ferrarese05rev,Graham08,Gultekin09}). Given the relevance of $M_{\rm BH}$ measurements, and the still small numbers available, it is essential to progress with the observations and models of new galaxies to populate the scaling relations. Together with the relations themselves, their intrinsic scatter is also important. It is used to test models of the joint evolution of BHs and galaxies and is required to estimate the mass function of supermassive black holes in the Universe \citep{Yu02,Marconi04,Lauer07}. For the latter one needs a reliable estimate of the measurements errors. Here we present an update on our efforts to address these issues: (i) Populate the BH scaling relations and (ii) test the accuracy of the $M_{\rm BH}$ measurements.

\section{Models and Results}

Our models are constrained by high spatial resolution (but seeing-limited) integral-field observations of the stellar kinematics obtained with \oasis\ \cite{mcdermid06}, in combination with large field \sauron\ kinematics \citep{emsellem04}. We adopt the Multi-Gaussian Expansion (MGE) parametrization of the surface brightness \cite{emsellem94}. We apply two independent axisymmetric modeling methods to fit the same stellar kinematics and MGEs: (i) An orbit-superposition Schwarzschild's \cite{schwarzschild79} approach as implemented in \cite{cappellari06} and (ii) the Jeans Anisotropic MGE (JAM) method \cite{Cappellari08}. The availability of good \oasis\ kinematics, Hubble Space Telescope (HST) imaging and the exclusion of objects clearly inconsistent with the assumed axisymmetry lead to a sample of 25 early-type galaxies.

\begin{figure}
  \includegraphics[width=\textwidth]{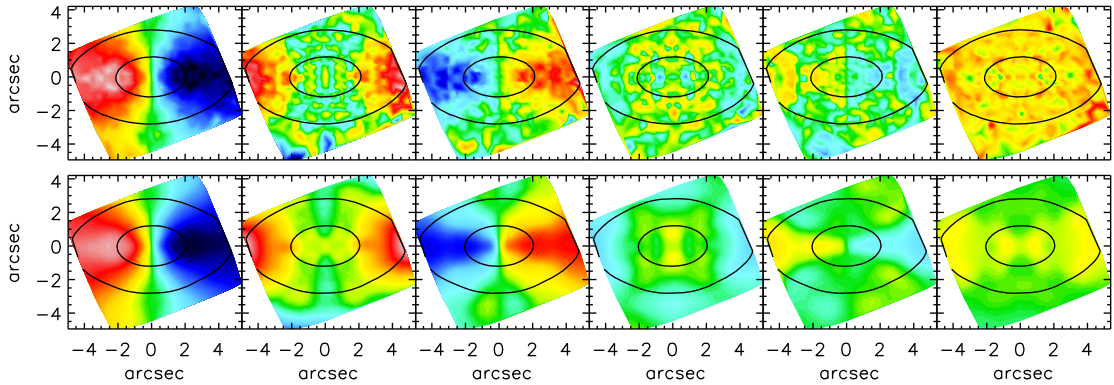}
  \caption{Data-model comparison for the Schwarzschild model. The bi-symmetrized \oasis\ stellar kinematics \cite{mcdermid06} of NGC~4473 (top panels) is compared to the best-fitting axisymmetric Schwarzschild model (bottom panels). From left to right the plots show the mean velocity $V$, the velocity dispersion $\sigma$ and the Gauss-Hermite moments $h_3$--$h_6$. Contours of the galaxy surface brightness are overlaid.}
  \label{fig:schw_fit}
\end{figure}

\begin{figure}
  \includegraphics[width=0.7\textwidth]{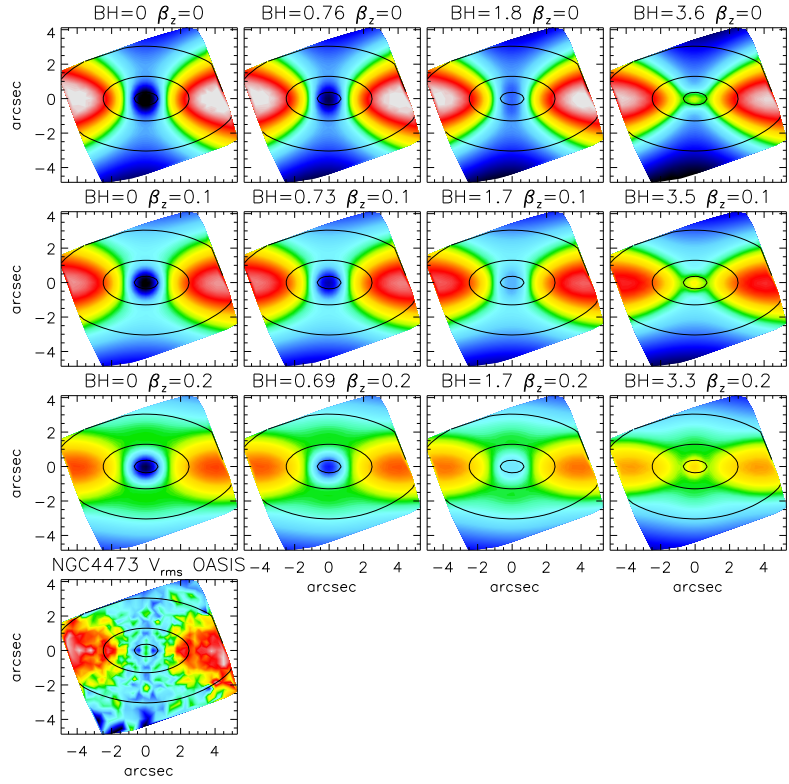}
  \caption{Data-model comparison for the JAM models. The first three rows show the $V_{\rm rms}$ of the JAM models for different $M_{\rm BH}$ (in units of $10^8$ M$_\odot$) and anisotropy $\beta_z=1-\sigma_z^2/\sigma_R^2$ (above each plot), with $M/L$ optimized to best fit the \oasis\ data. The bottom-left panel is the bi-symmetrized $V_{\rm rms}=\sqrt{V^2+\sigma^2}$ observed with \oasis. The best-fit can be done by eye and it agrees with the $\Delta\chi^2$ minimum of Fig.~\ref{fig:chi2}.}
  \label{fig:jam_fit}
\end{figure}

\begin{figure}
  \includegraphics[width=0.49\textwidth]{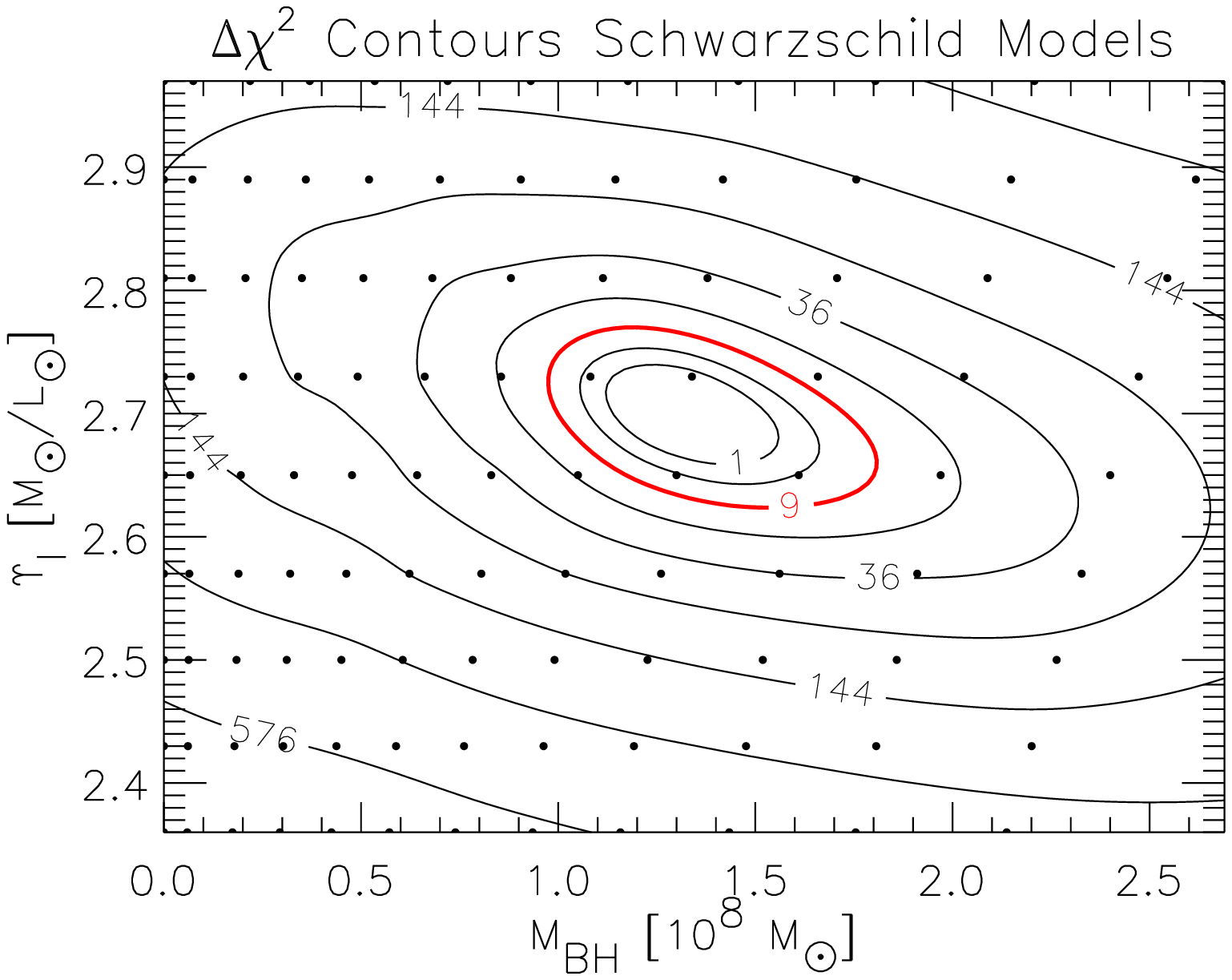}
  \includegraphics[width=0.50\textwidth]{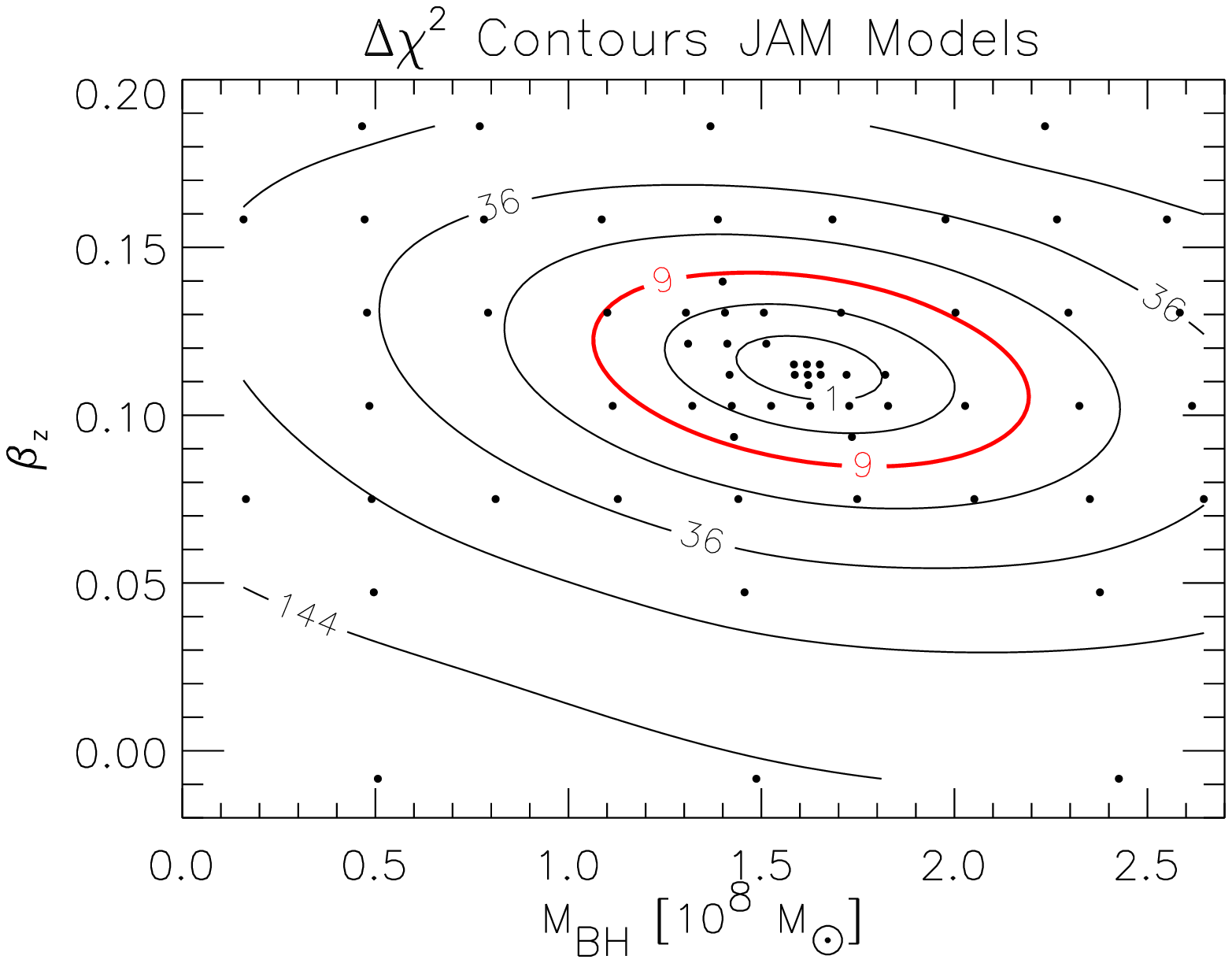}
  \caption{{\em Left Panel:} Contours of $\Delta\chi^2$ describing the agreement between the \oasis$+$\sauron\ data and the Schwarzschild dynamical models, as a function of the $M_{\rm BH}$ and $M/L$. {\em Right Panel:} Adaptively-sampled contours of $\Delta\chi^2$ between the \oasis\ data and the JAM models, as a function of $M_{\rm BH}$ and the anisotropy $\beta_z$. For each model the $M/L$ was optimized to best fit the \oasis\ data. The three lowest contours correspond to the 1, 2, and $3\sigma$ (thick line) confidence levels, marginalized for one parameter. The black dots show the models that were actually computed, while the contours were interpolated with a thin-plate spline.}
  \label{fig:chi2}
\end{figure}

An example of the results obtained with our two modeling approaches on NGC~4473 is shown in Fig.~\ref{fig:schw_fit}, \ref{fig:jam_fit} and \ref{fig:chi2}. This galaxy is an interesting case as it shows no peak in the central $V_{\rm rsm}$ (or $\sigma$) and one may wonder whether a BH is really present and what observable is actually constraining its mass. It was also modeled by \cite{gebhardt03} who derived $M_{\rm BH}=(1.3^{+0.5}_{-0.9})\times10^8$ M$_\odot$ (1$\sigma$ errors), using HST/STIS kinematics. The value is consistent with our \oasis\ determinations, but it has larger errors. Even though the radius of the BH sphere of influence is $R_{\rm BH}\approx0.24''$ while our seeing PSF has a 3.3$\times$ larger ${\rm FWHM}\approx0.80''$, the BH can be reliably measured. Fig.~\ref{fig:jam_fit} illustrates the fact that the BH detection in this object comes from the lack of the clear drop in $V_{\rm rsm}$ that one would expect to see if there was no BH, with our spatial resolution. The integral-field data are essential to constrain the anisotropy and provide a clear BH detection.

\begin{figure}
  \includegraphics[width=0.5\textwidth]{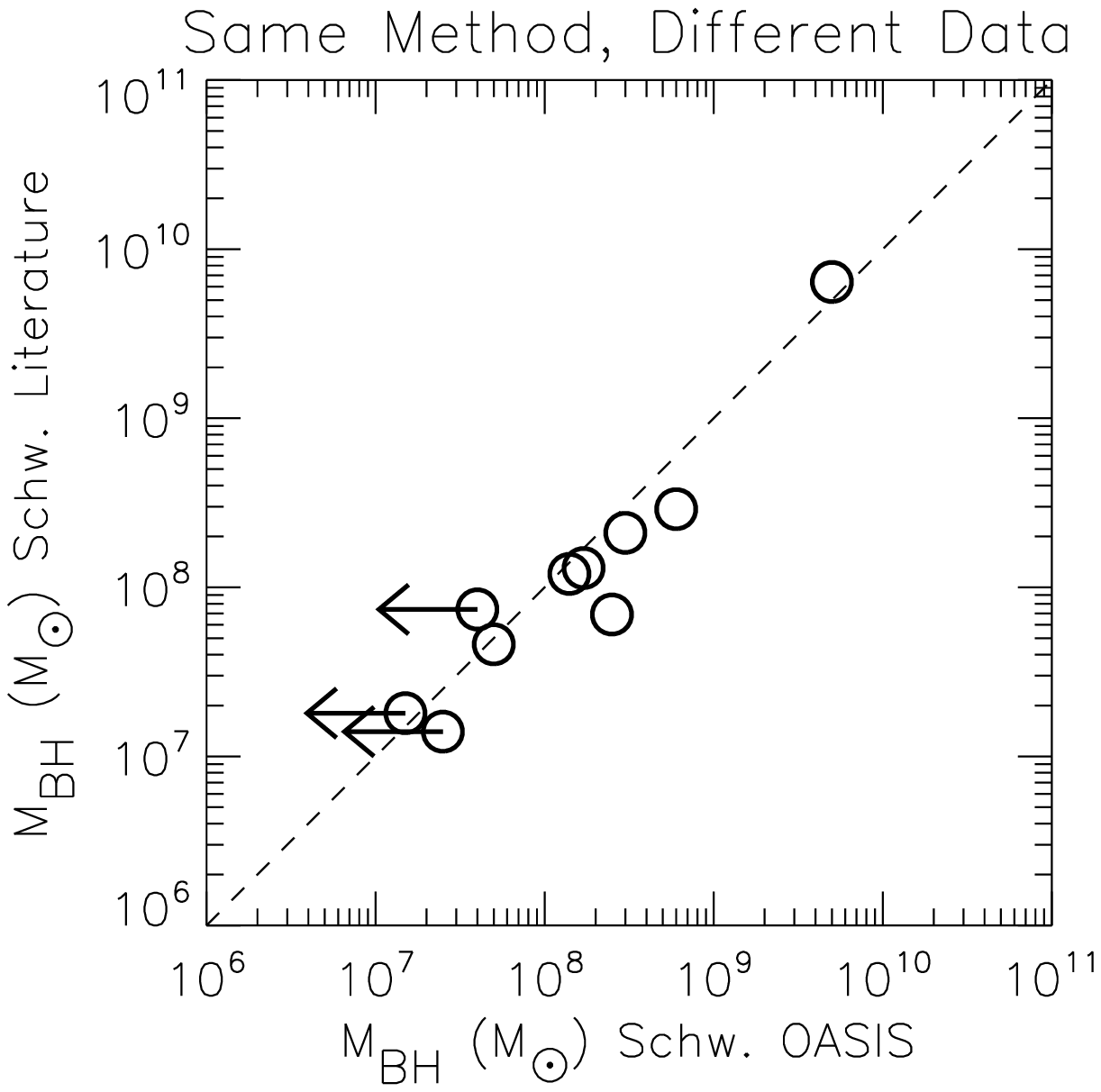}
  \includegraphics[width=0.5\textwidth]{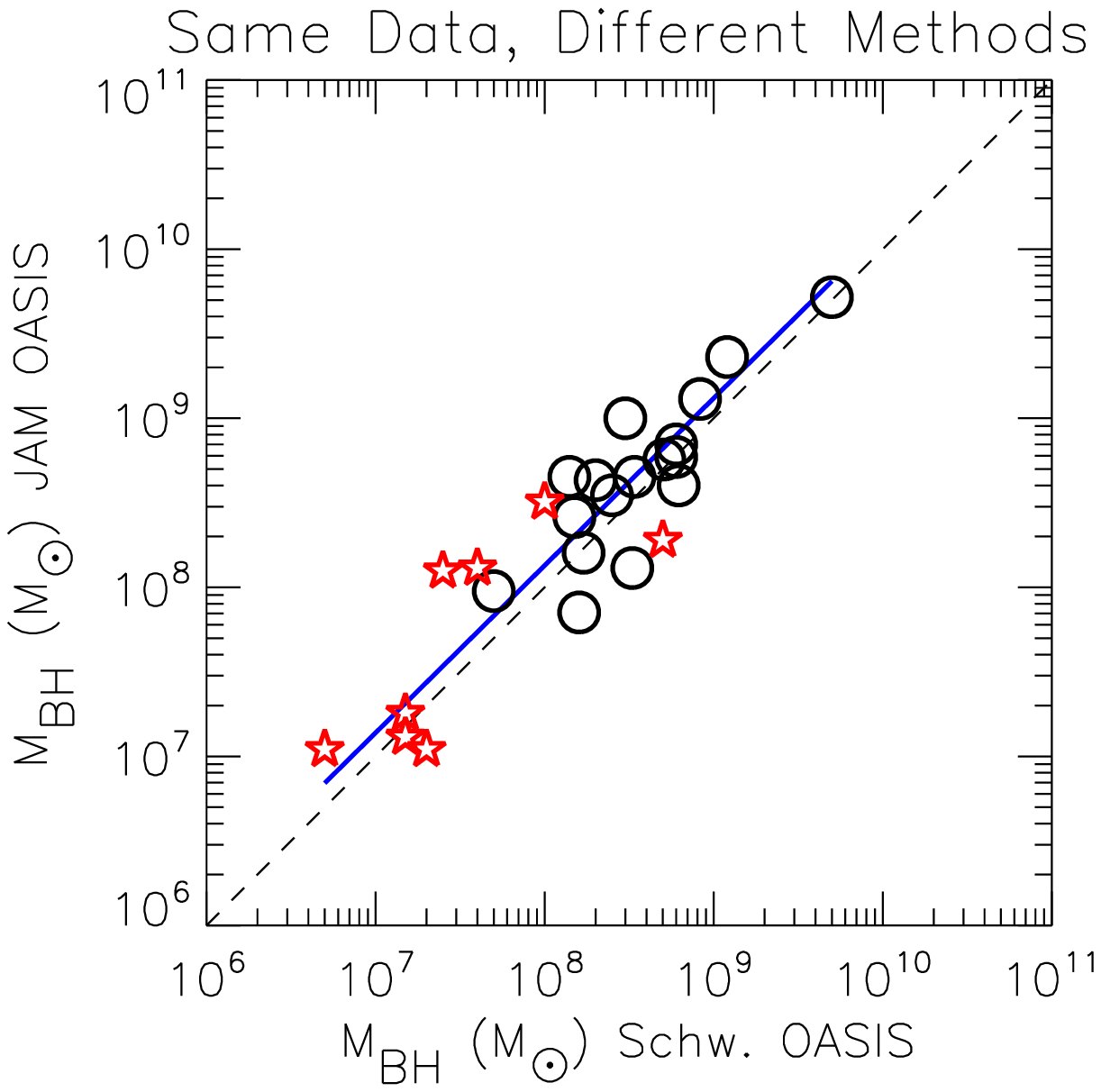}
  \caption{{\em Left Panel:} Comparison between the $M_{\rm BH}$ measured with our Schwarzschild models of the \oasis\ integral-field kinematics and published $M_{\rm BH}$ from \cite{Gultekin09} using HST/STIS kinematics. Upper limits in our models are indicated by an arrow. {\em Right Panel:} Comparison between our $M_{\rm BH}$ from Schwarzschild and from JAM models of the same \oasis\ data. Upper limits are indicated by star symbols.}
  \label{fig:comparison}
\end{figure}

In Fig.~\ref{fig:comparison} we confront our $M_{\rm BH}$ determinations, using Schwarzschild's method and \oasis\ data, to 10 measurements from the compilation of \cite{Gultekin09} using the same method but HST/STIS kinematics. There is a reassuring agreement between the two completely independent determinations. However the differences between the two sets of values would imply a $1\sigma$ error on $M_{\rm BH}$ of $\sim0.20$ dex, which is significantly larger than the typical formal errors. This is likely due to unaccounted systematics in the data and differences in the modeling implementations. We measure a similar scatter when comparing all our 25 $M_{\rm BH}$ values with Schwarzschild's method against the ones obtained with JAM. As the former method is more general than the latter, this scatter gives an upper limit to the errors in our \oasis\ determinations, however it does not take into account systematics in the data and in the modeling assumptions.
Below $M_{\rm BH}\lesssim10^8$ M$_\odot$ most of our \oasis\ measurements are upper limits. The low $M_{\rm BH}$ regime must be explored with observations using adaptive optics as illustrated in this volume by \cite{Krajnovic10,Seth10}.

\end{document}